\documentstyle[prd,aps]{revtex}
\begin{document}
\preprint{PU-RCG/98-15v3, gr-qc/9809062}
\draft
\renewcommand{\topfraction}{0.8}
\renewcommand{\bottomfraction}{0.8}
\twocolumn[\hsize\textwidth\columnwidth\hsize\csname 
@twocolumnfalse\endcsname
\title{Duality Invariance of Cosmological Perturbation Spectra} 
\author{David Wands}
\address{School of Computer Science and Mathematics, University of 
Portsmouth, Portsmouth PO1 2EG, U.~K.}  
%\date{\today} 
\maketitle
\begin{abstract}
I show that cosmological perturbation spectra produced from quantum
fluctuations in massless or self-interacting scalar fields during an
inflationary era remain invariant under a two parameter family of
transformations of the homogeneous background fields. This relates
slow-roll inflation models to solutions which may be far from the
usual slow-roll limit.  For example, a scale-invariant spectrum of
perturbations in a minimally coupled, massless field can be produced
by an exponential expansion with $a\propto e^{Ht}$, or by a
collapsing universe with $a\propto (-t)^{2/3}$. 
\end{abstract}
\pacs{{\em to appear in Phys Rev D}\ \ \ PACS numbers: 98.80.Cq \hfill
Preprint PU-RCG/98-15v3, gr-qc/9809062}
\vskip2pc]
 
%%%%%%%%%%%%%%%%%%%%%%%%%%%%%%%%%%%%%%%%%%%%%%%%%%%%%%%%%
 
%\section{Introduction}

The spectrum of perturbations on large scales is a key test of any
models of the early universe.  During an inflationary era quantum
fluctuations on small scales become stretched beyond the horizon
generating inhomogeneities on super-horizon scales that are otherwise
inexplicable in the standard big bang model~\cite{gw,deltaphi,BST}.
Conventional models of slow-roll inflation generally predict an almost
scale-invariant spectrum of adiabatic density
perturbations~\cite{LL93}.  As major observational programs are now
underway to produce detailed maps of these perturbations, it is
important to investigate whether one can uniquely reconstruct the
inflationary history of the universe from the spectrum of
inhomogeneities.  This question has received considerable attention in
recent years in the context of slow-roll inflation~\cite{recon}
where it has been realised that there is a degeneracy in the spectrum
of adiabatic density perturbations.  This could be removed by a
detection of the gravitational wave background on the same scale
which, in the slow-roll approximation, gives a direct record of the
evolution of the scale factor, and hence of the inflaton potential.

By contrast there has been relatively little study of the
reconstruction of the evolution in the non-slow-roll
case~\cite{nonSR}. Recently it has been discovered that the spectra of
perturbations produced in so-called pre-big bang models of the early
universe~\cite{pbb}, based on solutions of the low energy string
effective action, are invariant under SL(2,R) symmetry transformations
(including S-duality transformations) of the background
fields~\cite{CEW97}. This raises the interesting question of what is
the most general type of cosmological evolution that yields a given
perturbation spectrum.

I will consider linear perturbations, $\delta\phi(\eta,x^i)$, 
about a homogeneous background, $\phi(\eta)$, in a homogeneous cosmology.
For a minimally coupled massless field we can neglect any back-reaction 
upon the spacetime curvature to first order, and so perturbations obey the
wave equation
\begin{equation}
\label{eom}
\ddot{\delta\phi} + 3H\dot{\delta\phi} - \nabla^2\delta\phi = 0
\ ,
\end{equation}
where a dot denotes derivatives with respect to cosmic time $t$, and
$3H$ corresponds to the expansion rate of the homogeneous 
hypersurfaces.
One can decompose the perturbations into independent wavemodes 
$\delta\phi_k(\eta)Q_k(x^i)$, where $Q_k(x^i)$ is the eigenfunction 
of the spatial Laplacian $\nabla^2$ with eigenvalue $-k^2$.

The canonically normalised quantum field corresponds to the
conformal field perturbation $u=a\delta\phi$, where the scale factor
$a=\int Hdt$.  Perturbations obey the wave equation
\begin{equation}
\label{ukeom}
u_k'' + ( k^2 + \mu^2 ) u_k = 0 
 \ ,
\end{equation}
which corresponds to an oscillator with a time dependent effective 
mass-squared
\[
\mu^2= - {a'' \over a} \ ,
\]
%where $h=aH$ is the comoving Hubble rate, and 
where a prime denotes a
derivative with respect to conformal time $\eta=\int dt/a$.
During a conventional inflationary era $\mu^2$ is negative and decreases
monotonically, leading to the amplification of vacuum fluctuations.
Modes on arbitrarily small scales ($k^2/|\mu^2|\to\infty$) are
presumed to occupy the flat spacetime vacuum state ($u_k\to
e^{-ik\eta}/\sqrt{2k}$).
These vacuum fluctuations eventually lead to a specific spectrum
of perturbations on large scales ($k^2/\mu^2\to0$), the form of
which is determined solely by $\mu^2(\eta)$.

Consider the most commonly studied case of a power-law
expansion~\cite{PLI}, where the scale factor grows as $a\propto t^p$,
which corresponds, in terms of the conformal time, to
\begin{equation}
\label{pla}
a = a_0 \left( {\eta \over \eta_0} \right)^{(1-2\nu)/2} \ ,
\end{equation}
where 
\begin{equation}
\nu = {3\over2} + {1\over p-1} \ .
\end{equation}
Note that $a\to\infty$ as $t\to\infty$ only coincides with the limit
$\eta\to\infty$ for $p<1$. During an inflationary expansion with $p>1$
there is an event horizon and $\eta\to0$ from below as $a\to\infty$.

The effective mass in Eq.~(\ref{ukeom}) is
\begin{equation}
\mu^2=-{\nu^2-1/4 \over \eta^2} \ .
\end{equation} 
Note that for $\nu=\pm1/2$ the effective mass $\mu^2$ vanishes and
there is no particle production. This corresponds to a static universe
($p=0$) or a spatially flat Friedmann-Robertson-Walker (FRW) radiation
dominated universe ($p=1/2$).
During an inflationary expansion with $p>1$ and $\nu>3/2$,
$\mu^2\to-\infty$ as $\eta\to0^-$ which leads to fluctuations on
scales larger than the horizon, $|k\eta|>1$.
However, it has previously been noted~\cite{deflation} that a
collapsing universe could in principle produce large scale
perturbations from small scale quantum fluctuations. $|\eta|$ also
represents an effective event horizon in a collapsing model with $p<1$
where $\eta\to0$ from below as $a\to0$.
Any given comoving mode $k$ gets pushed outside horizon
as $|\eta|$ decreases and $k\eta\to0$ as $t\to0^-$ for $p<1$.
For $|\nu|<1/2$, i.e., $0<p<1/2$, the effective mass-squared $\mu^2$ is
positive and becomes large as $\eta\to0^-$ which strongly suppresses
fluctuations on large scales ($|k\eta|\ll1$).

The general solution of the wave equation~(\ref{ukeom}) during a
power-law expansion/contraction is~\cite{LS}
\begin{equation}
u_k(\eta) = \sqrt{|k\eta|} \left[ u_+ H_{|\nu|}^{(1)}(|k\eta|) 
+ u_- H_{|\nu|}^{(2)}(|k\eta|) \right] \ . 
\end{equation}
where $H^{(i)}_{|\nu|}$ are Hankel functions of order $|\nu|$.
Choosing the quantum vacuum state at early times on small scales 
($k\eta\to-\infty$) then determines the spectrum of perturbations
on large scales ($k\eta\to0$)
\begin{equation}
\label{Pu}
{\cal P}_u = {C^2(|\nu|) k^2 (-k\eta)^{1-2|\nu|} \over (2\pi)^2} \ ,
\end{equation}
where the power spectrum is conventionally 
defined as ${\cal P}_u=k^3|u_k^2|/2\pi^2$, and the numerical coefficient
\begin{equation}
C(|\nu|) \equiv {2^{|\nu|}\Gamma(|\nu|) \over 2^{3/2}\Gamma(3/2)} \ .
\end{equation}
The spectrum of scalar field perturbations produced on large scales
($|k\eta|\gg1$) can therefore be written as
\begin{equation}
\label{Pdphi}
{\cal P}_{\delta\phi} 
= \left( {C(|\nu|) \over \nu-1/2} \right)^2
 \left( {H \over 2\pi} \right)^2 (-k\eta)^{3-2|\nu|} \ .
\end{equation}
In the limit of de Sitter expansion in flat FRW, $p\to\infty$ and
$\nu\to3/2$, we recover the famous result ${\cal
P}_{\delta\phi}=(H/2\pi)^2$ at horizon crossing ($k\eta=-1$) and the
spectrum is independent of scale.

Notice, however, that the spectrum ${\cal P}_u$ given in
Eq.~(\ref{Pu}) is invariant under the transformation
$\nu\to\tilde\nu=-\nu$ or, equivalently,
\begin{equation}
\label{ptop}
p \to \tilde{p} = {1-2p \over 2-3p} \ .
\end{equation}
The perturbation spectrum produced during a power-law inflationary
expansion with $p>1$ is indistinguishable from the spectrum produced
during a power-law collapse $\tilde{p}<1$, where $\tilde{p}$ is given
by Eq.~(\ref{ptop}). There are two fixed points where
$\tilde{p}=p$. These occur where $p=1/3$ or $p=1$, which correspond to
$\nu=0$ and $|\nu|\to\infty$ respectively.

Thus one obtains a scale invariant spectrum of perturbations not
just for de Sitter inflation in flat FRW (where 
$p\to\infty$) but also for $p=2/3$, which corresponds to a collapsing
dust-dominated FRW universe.  This result is rather surprising at first
sight since the scale invariance of the de Sitter spectrum can be
understood as being due to the time-invariant nature of this solution, and a
collapsing dust universe seems to be far from static. However, there
is an important difference between the two cases. For $\nu>0$ (which
includes conventional inflation models with $p>1$) the scalar field
perturbations become frozen in on large scales as $H^2\propto\eta^{2\nu-3}$ in
Eq.~(\ref{Pdphi}). Thus ${\cal P}_{\delta\phi}$ remains constant for
any given mode $k$ as $|k\eta|\to0$. But for $\nu<0$, the
perturbations grow outside the horizon with ${\cal
P}_{\delta\phi}\propto \eta^{-4|\nu|}$ as $|k\eta|\to0$. The amplitude
of the perturbations as they cross outside the horizon ($|k\eta|=1$)
grows as $H^2\propto\eta^{2\nu-3}$, and thus the amplitude of modes already
outside the horizon grows at precisely the same rate for $\nu=-3/2$
and at any given time the spectrum is scale-invariant on
large-scales ($|k\eta|\ll1$).

If one asks what is the most general cosmological evolution which will
lead to an equivalent time-dependent mass for the perturbations and a
scale-invariant spectrum of perturbations in massless fields, one
obtains the simple solution
\begin{equation}
\label{tildeHZ}
\tilde{a}(\eta) = C_1
 \left[ \left({\eta\over\eta_1}\right)^{-1}
 + \left({\eta\over\eta_1}\right)^2 \right] \ ,
\end{equation}
which describes a non-singular metric smoothly interpolating between a
collapsing dust solution at early times ($\eta\to-\infty$) and an
exponentially expanding de Sitter solution at late times ($\eta\to0$).

One can go on to ask whether given any particular cosmological
solution $a(\eta)$ one can write down the most general evolution
$\tilde{a}(\eta)$ that would give rise to an equivalent time-dependent
mass-squared, $\mu^2$, and hence perturbation spectrum ${\cal P}_u$.
The answer turns out to be that the same spectrum of perturbations on
large scales will be produced by the two parameter family of solutions
\begin{equation}
\label{atoa}
a(\eta) \to \tilde{a}(\eta)
 = C a(\eta) \int_{\eta_*}^\eta {d\eta' \over a^2(\eta')} \ .
\end{equation}
The constant $C$ describes an arbitrary rescaling of the whole metric
which does not change the essential physics of the solutions, but the
constant of integration $\eta_*$ describes a one parameter family of
different solutions. 

For example, substituting in the power-law inflationary
solutions given in Eq.~(\ref{pla}) one obtains
\begin{equation}
\label{pltildea}
\tilde{a}(\eta) = C_1 \left({\eta\over\eta_1}\right)^{1/2}
 \left[ \left({\eta\over\eta_1}\right)^{\nu}
 + \left({\eta\over\eta_1}\right)^{-\nu} \right] \ .
\end{equation}

Gravitational waves (transverse, traceless perturbations of the
metric) in Einstein gravity obey the same wave equation as a minimally
coupled massless scalar field~\cite{gw} and hence the graviton
spectrum is proportional to ${\cal P}_{\delta\phi}$. This is often
assumed to give an unambiguous record of the evolution of the
cosmological scale factor, $a(\eta)$. However an identical spectrum of
gravitational waves will be produced by the two parameter family of
solutions given in Eq.~(\ref{atoa}).

This invariance of cosmological perturbation spectra has already been
noted in the context of superstring cosmology where the perturbation
spectra of fields in the low energy string effective action may be
invariant under symmetries of the action. In the pre big bang
scenario~\cite{pbb} the graviton and dilaton fields are minimally
coupled in the conformal Einstein frame where the metric evolves as
$a\propto t^{1/3}$. $p=1/3$ is a fixed point of the transformation
given in Eq.~(\ref{ptop}) and the graviton and dilaton spectra on
large scales remain invariant under T-duality or S-duality
transformations of the background model. However the axion-type fields
are minimally coupled in the conformally related axion
frames~\cite{CEW97,CLW98}. SL(2,Z) S-duality transformations of the
power-law vacuum solutions lead to a scale factor in the axion frame
which evolves as given in Eq.~(\ref{pltildea})~\cite{CEW97}. By
constructing explicitly SL(2,R) invariant perturbation variables it
was shown that both the axion and dilaton spectra remained invariant
under arbitrary SL(2,R)
transformations~\cite{CEW97}. Equation~(\ref{atoa}) generalises this
result to arbitrary background solutions for $a(\eta)$, and to
theories which may or may not have their origin in superstring theory.

The wave equation~(\ref{ukeom}) for the perturbation $u$ may be
derived from an effective action
\begin{equation}
\label{myS}
S = {1\over2} \int d\eta \int d^3x \left\{ u'^2 - u_{,i}u_{,i} - \mu^2
u^2 \right\} \ ,
\end{equation}
with the corresponding Hamiltonian
\begin{equation}
\label{myH}
{\cal H} = {1\over2} \int d^3x \left\{ \pi_u^2 +
 u_{,i}u_{,i} + \mu^2 u^2 \right\} \ ,
\end{equation}
where the momentum canonically conjugate to $u$ is $\pi_u=u'$. 
The action $S$ and Hamiltonian ${\cal H}$ both remain invariant under
the transformation given in Eq.~(\ref{atoa}) which leaves $u(\eta)$
and $\mu^2(\eta)$ invariant.
It is interesting to compare this with a different invariance which has
also recently been noted in the context of superstring
cosmology~\cite{BMUV}, and applied to generalised cosmological
perturbations~\cite{BGV}. This is an invariance of the effective
action
\begin{equation}
\label{theirS}
\hat{S} = {1\over2} \int d\eta \int d^3x \, a^2 \left\{ \delta\phi'^2 -
 \delta\phi_{,i}\delta\phi_{,i} \right\} \ ,
\end{equation}
and corresponding Hamiltonian
\begin{equation}
\label{theirH}
\hat{{\cal H}} = {1\over2} \int d^3x \left\{ a^{-2}
 \pi_{\delta\phi}^2 + a^2 \delta\phi_{,i}\delta\phi_{,i} \right\} \ ,
\end{equation}
written in terms of the field perturbation $\delta\phi$ and its
conjugate momentum $\pi_{\delta\phi}=a^2\delta\phi'$.  
The Lagrangian in Eq.~(\ref{theirS}) differs from that in
Eq.~(\ref{myS}) by a total derivative
\begin{equation}
\hat{S} = S - {1\over2} \int d\eta \int d^3x \, {d \over d\eta} \left(
{a'\over a} u^2 \right) \ ,
\end{equation}
which does not affect the equation of motion, Eq.~(\ref{eom}), but does
change the Hamiltonian
\begin{equation}
\hat{{\cal H}} = {\cal H} + \int d^3x \left\{ {a''\over a}u^2 - {1\over2}
\left( {a'\over a} u^2 \right)^{\prime} \right\} \ . 
\end{equation}
There is a duality invariance of the action $\hat{S}$ and Hamiltonian
$\hat{{\cal H}}$ under which the ``pump field''
$a^2\to\tilde{a}^2=a^{-2}$ is inverted and the field perturbation
$\delta\phi$ is exchanged with its canonical momentum
$\pi_{\delta\phi}$~\cite{BGV}.
The Hamiltonian $\hat{{\cal H}}$ does not remain invariant under the
transformation in Eq.~(\ref{atoa}), but neither does the Hamiltonian
${\cal H}$ remain invariant under the duality transformation in
Ref.\cite{BGV}. The effective action is only defined up to boundary
terms and due to the explicit time-dependence of $a(\eta)$, the
Hamiltonian is not uniquely defined. Both transformations, however, represent
symmetries of the equation of motion.

The most cosmologically significant perturbation spectrum
produced during an inflationary era in the early universe is likely to
be the primordial spectrum of adiabatic density perturbations on large
scales induced by the perturbations in the scalar field which drives
inflation. To study the evolution of this field requires us to include
the self-interaction potential of the field and the back-reaction of
metric fluctuations. Fortunately Mukhanov~\cite{Mukhanov} has shown
that the wave equation for the gauge invariant field perturbation
\begin{equation}
u = a \left[ \delta\phi + \dot\phi {\psi \over H} \right] \ ,
\end{equation}
where $\psi$ is the gauge-dependent curvature perturbation~\cite{MFB},
can still be written in the form given in Eq.~(\ref{ukeom}) but with a
time-dependent mass-squared
\begin{equation}
\label{muz}
\mu^2 = -{z'' \over z} \ ,
\end{equation}
where $z=a\dot\phi/H$. 
Quite generally we can write
\begin{equation}
\label{zgamma}
z = a \sqrt{ {3\gamma \over 8\pi G} } \ ,
\end{equation}
where the effective barotropic index
$\gamma\equiv\dot\phi^2/(V+\dot\phi^2/2)$. In the special case of
power-law inflation driven by a scalar field with exponential
potential, $\dot\phi\propto H$ and hence $\gamma$ is a constant and we
have $z\propto a$.

Starting from any known solution $z(\eta)$ we obtain the identical
spectrum of perturbations ${\cal P}_u$ from the two parameter family
of solutions
\begin{equation}
\label{ztoz}
z(\eta) \to \tilde{z}(\eta)
 = C z(\eta) \int_{\eta_*}^\eta {d\eta' \over z^2(\eta')} \ .
\end{equation}
which leaves $\mu^2$ given in Eq.~(\ref{muz}) invariant.

The gauge-invariant curvature perturbation $\zeta$~\cite{BST,MFB,LL93}
is related to the field perturbation $u$ by
\begin{equation}
\zeta = \psi + {H\over\dot\phi} \delta\phi = {u \over z} \ .
\end{equation}
This is usually evaluated in terms of the quantities at horizon
crossing. This is because $\zeta$ becomes constant on super-horizon
scales for adiabatic perturbations. In this case the $z$ acquires an
implicit scale dependence due to the different times at which
different scales are evaluated. However one can also evaluate $\zeta$
at a fixed time, such as the end of inflation, in which case the scale
dependence of $\zeta$ is due solely to the scale dependence of $u$,
and $z$ contributes a scale independent factor. 
Thus under the transformation given by Eq.~(\ref{ztoz}) the curvature
perturbation is rescaled by an overall factor $z/\tilde{z}$, but the
spectral index
\begin{equation}
n \equiv 1 + {d\ln {\cal P}_\zeta \over d\ln k} \ ,
\end{equation}
remains invariant. 

For instance, it is well-known that the extreme slow-roll limit of
inflation corresponding de Sitter expansion driven by a massless
scalar field, where $z\propto \eta^{-1}$, leads to a scale-invariant
Harrison-Zel'dovich ($n=1$) spectrum of curvature
perturbations. However substituting this familiar form for $z(\eta)$
into Eq.~(\ref{ztoz}) yields the most general evolution which gives a
scale-invariant spectrum as
\begin{equation}
\label{altHZ}
\tilde{z}(\eta) = C_1
 \left[ \left({\eta\over\eta_1}\right)^{-1}
 + \left({\eta\over\eta_1}\right)^2 \right] \ .
\end{equation}
This shows that it is in fact possible to produce a scale-invariant
spectrum of curvature perturbations from inflation that is far from
the usual slow-roll limit. 

Unfortunately it is not possible to uniquely determine the form of the
self-interaction potential $V(\phi)$ for a given $z(\eta)$, such as
that given in Eq.~(\ref{altHZ}). For example, both power-law
inflation~\cite{PLI} driven by an exponential potential, and natural
inflation~\cite{natural} where the potential energy remains
effectively constant, can give rise to a power-law spectrum of
curvature perturbations with $n=$constant~\cite{SL}.  However it is
possible to test the consistency of the slow-roll approximation for a
given $z(\eta)$.  The slow-roll approximation requires that the
effective barotropic index, $\gamma$ in Eq.~(\ref{zgamma}), is small
and slowly varying, so that to zeroth order in the slow-roll
parameters~\cite{recon}, the evolution of $z$ is determined by the
growth of the scale factor $a\sim\eta^{-1}$. This implies that
$z''z/z^{\prime2}\approx2$.  For the general form of
$\tilde{z}(\eta)$ which yields a scale-invariant spectrum of curvature
perturbations, given in Eq.~(\ref{altHZ}), this slow-roll condition is
badly broken at early times for $|\eta/\eta_1|\gg1$.

Even in the slow-roll limit, the spectrum of curvature perturbations
is sufficient only to determine the inflaton potential up to a one
parameter class of solutions~\cite{recon}. The amplitude of the
gravitational wave perturbations is then required to fix the actual
amplitude of the inflaton potential. In this paper I have demonstrated
that if one allows behaviour which may be far from the slow-roll limit
there is a degeneracy even in the spectrum of gravitational wave
perturbations.  The general solution which yields an almost
scale-invariant spectrum of gravitational waves interpolates between
an initially collapsing universe and a quasi-de Sitter
expansion. However, the asymptotic behaviour at late times reproduces
the usual slow-roll result, so in practice this need not be a serious
limitation for reconstructing the evolution in the context of
conventional inflation models~\cite{recon}.

On the other hand the transformation presented here suggests that it
might be possible to use slow-roll techniques to analyse perturbations
in models far from the usual slow-roll limit if they can be related to
slow-roll models.  An example of this is provided by solutions to the
low energy string effective action where perturbation spectra in
general axion-dilaton cosmologies can be related to much simpler
dilaton-vacuum solutions by a duality transformation~\cite{CEW97}.

\acknowledgements

DW would like to thank John Barrow, Ed Copeland, Andrew Liddle, Jim
Lidsey and Karim Malik for helpful comments.

\end{document}